\documentclass[preprint,5p]{elsarticle}

\usepackage{amsmath}
\usepackage{amssymb}

\newcommand{\beq}{\begin{eqnarray}}
\newcommand{\eeq}{\end{eqnarray}}

\newcommand{\cm}{{\rm cm}}
\newcommand{\pc}{{\rm pc}}
\newcommand{\s}{{\rm s}}

\newcommand{\MeV}{{\rm MeV}}
\newcommand{\GeV}{{\rm GeV}}
\newcommand{\mpl}{m_{\rm Pl}}

\begin{document}

\begin{frontmatter}

\title{Clusters of Black Holes as Point-Like Gamma-ray Sources}

\author[mephi]{K.~M.~Belotsky}
\ead{k-belotsky@yandex.ru}

\author[mephi]{A.~V.~Berkov}
\ead{berkov@migmail.ru}

\author[mephi]{A.~A.~Kirillov}
\ead{kirillov-aa@yandex.ru}

\author[mephi]{S.~G.~Rubin}
\ead{sergeirubin@list.ru}

\address[mephi]{National Research Nuclear University MEPhI, 31, Kashirskoe shosse, Moscow, Russia}

\begin{abstract}
	The possibility of identifying some of Galactic gamma-ray  sources as clusters of primordial black holes is discussed. The known scenarios of supermassive black hole formation indicate the multiple formation of lower-mass black holes. Our analysis demonstrates that due to Hawking evaporation the cluster of black holes with masses about $10^{15}$~g could be observed as a gamma-ray source. The total mass of typical cluster is $\sim 10 M_\odot$. Detailed calculations have been performed on the basis of specific model of primordial black hole formation.
\end{abstract}


\end{frontmatter}

\section{Introduction}\label{sec:intro}

Among numerous unsolved problems of astrophysics, there are two seemingly independent problems: first, the formation of
supermassive black holes in galactic nuclei \cite{bib:Dolgov1, bib:Dokuchaev1, bib:Dolgov2, bib:Rees, bib:Begelman, bib:Dokuchaev2, bib:Dokuchaev3, bib:Dokuchaev4} and, second, the existence of unidentified gamma-ray sources observed by the Compton Gamma-Ray Observatory (CGRO) \cite{bib:3EG, bib:EGR, bib:Thompson}. The Energetic Gamma Ray Experiment Telescope (EGRET) operating in the energy range from 30~MeV to 30~GeV \cite{bib:Thompson, bib:EGRET.Calibration} is the main instrument of this observatory. The existence of such sources has also been confirmed on the Fermi Gamma-ray Space Telescope FERMI \cite{bib:1FGL} equipped with the Large Area Telescope (LAT) on board, which covers the energy range from 20~MeV to 300~GeV \cite{bib:FERMI.Mission}.

The total view of the celestial sphere in the gamma-ray range was obtained for the first time on the CGRO (1991--2000) with the EGRET; this view revealed a surprisingly large number of unidentified point-like gamma-ray sources (170 of 271), according to the 3EG catalog of gamma-ray sources \cite{bib:3EG}. However, owing to a certain improvement of the diffusion radiation model
used to analyze the observational data, the 3EG data were revised and the new EGR catalog of the sources was composed
\cite{bib:EGR}, where the number of unidentified sources is 87 of 188 and the distribution of the sources over the celestial sphere is more isotropic.

The Fermi Gamma-ray Space Telescope FERMI was successfully
launched into the geostationary orbit in summer 2008; the results obtained on this telescope were presented in February 2010 in the first FERMI 1FGL catalog \cite{bib:1FGL}. In contrast to the 3EG and EGR catalogs, the 1FGL catalog contains 630 unidentified sources of 1451. Their distribution on the celestial sphere is much more isotropic that those in the 3EG and EGR catalogs.

The second problem of modern astrophysics is the origin of
supermassive black holes (SMBH). Various formation scenarios of massive primordial black holes were discussed in
\cite{bib:Dolgov1, bib:Dokuchaev1, bib:Carr, bib:Rubin1}. Below we
briefly enumerate list the proposed mechanisms of primordial black
holes (PBHs) formation (see also the detailed review \cite{Carr}).

PBHs were first suggested as a result of adiabatic fluctuations at
the radiation-dominated stage of evolution of the Universe
\cite{carr75,zeldnov66,hawking71,Khlopov80}. Masses of such black hole do not exceed solar mass.

Variety of models with flat or bumped spectra of isocurvature
fluctuations were proposed
\cite{Barrow81,Bond82,Fuk86,Dolg87,Kof87,Kof88,Dol91,CDol92,Iok91}.
A very promising mechanism of SMBH formation from the large
amplitude isothermal fluctuations in baryonic charge density was
proposed by Dolgov and Silk \cite{DolgovSilk93}. In this mechanism
the bumped spectrum of isothermal fluctuations is a byproduct of
vacuum bubble collapses during phase transition at the
inflationary stage. Masses of black holes could be larger the
solar mass in orders of magnitude in this case.

Spectrum of fluctuations depends on the form of the inflaton
potential, see e.~g. \cite{Star92,ivan94}. As was shown in
\cite{Yoko}, it opens possibility to produce high density
fluctuations that collapse afterwards into SMBHs. In this case,
masses of black holes could be much larger than the solar mass.

Black holes with masses $\sim {\rm M}_{\odot}$ possibly formed at
quark-hadron phase transition at the cosmological time
$10^{-6}$~s, \cite{Jedamzik,Jedamzik2}. Today such black holes
would be a component of dark matter .

Suppose that some massive non-relativistic particles dominates at
an early stage of the Universe evolution. In that case the
pressure is negligible and could not prevent gravitational
collapse of high density regions \cite{Khlopov80}.

Usually, false vacuum decay is accompanied by formation of
spherical walls with a true vacuum inside
\cite{Bubble,Bubble2,Dodelson,Bubble3,Bubble4,Bubble5,Bubble6,Bubble7,Ru25}.
The walls quickly expand and collide what could lead to collapse
of islands of false vacuum into black hole. Most massive black
holes of order $1{\rm M}_{\odot}$ are produced during the period
of quark-hadron phase transition.

The mechanism of black holes formation from the closed domain
walls was proposed in \cite{bib:Rubin1,Ru01b} and developed in
\cite{Ru05}. These domain walls could be originated due to
evolution of a scalar field during inflation. An initial
non-equilibrium distribution of scalar field imposed by the
background de-Sitter fluctuations gives rise to the spectrum of
PBHs, which covers a wide range of mass --- from small masses up
to supermassive ones. The PBHs of smaller masses are concentrated
around the most massive ones forming a fractal-like cluster. It
was revealed that this mechanism is a rather common for many
inflationary models and it is worth to discuss it.

The  cosmological models of the formation of supermassive
primordial black holes can hardly be free of the multiple
formation of less massive black holes (see, e.g.,
\cite{bib:Rubin1,GraCo}). Moreover, to suppress the predicted abundance
of black holes of smaller mass, special assumptions are
required \cite{Ru10}. We devote special attention to the model of
PBH formation in the next sections.

The detection of black holes with masses below $10^4 M_{\odot}$ is
difficult in view of the weak accretion of surrounding matter.
Another method for detecting black holes by Hawking radiation
\cite{bib:Hawking} is efficient only in the vicinity of the Earth. Indeed, even in case of homogeneous distribution of PBHs over Galaxy, the rate of single PBH observation would be $\sim 10^{-6}$ years$^{-1}$. Meantime clusters containing a large number of black holes with masses below or about $10^{15}$g are visible from larger distances.

In this work, we demonstrate that clusters of primordial black holes can be detected as point-like gamma sources due to effect of Hawking evaporation. According to our estimations such a cluster contains black holes with different masses starting from tens of solar mass and smaller. Number of small black holes appears to be large enough so that their total radiation could be seen from the
Earth. As a result, the origin of unidentified gamma-ray sources can be explained and the way for detecting massive primordial black holes is indicated.

\section{Clusters of black holes from closed domain walls}

In this Section we briefly discuss specific model of massive black
hole formation. The details can be found in papers
\cite{Ru01b,Ru05} where the mechanism of massive primordial black
holes production was elaborated.

Some inflationary models suppose a creation of our Universe either
near a maximum of potential of inflaton field or near its saddle
point(s) to realize a desired slow rolling providing a sufficient
number of e-folds (see details, e.~g. in \cite{Dvali,Racetrack}).
As it will be shown below these models include the possibility of
the formation of macroscopically large closed walls from a scalar
field. After the end of inflation these closed walls collapse to
BHs if these walls are large and heavy enough \cite{bib:Rubin1,Ru01b}.
This mechanism is realized in well known models like the Hybrid
Inflation \cite{LindeHyb} and the Natural Inflation
\cite{Dolgov97}. A scalar field could be the inflaton itself or
some additional field.

Consider general mechanism of closed wall formation based on
quantum fluctuations near unstable point(s) like a saddle point or
a maximum of potential of scalar field. An evolving scalar field
may be split into classical part, governed by the classical
equation of motion, and quantum fluctuations \cite{Star}. To
facilitate the analysis, let us approximate the potential near its
maximum as
\begin{equation}
 V = V_{0} - \frac{m^{2}}{2} \phi^{2},
 \label{Vapprox}
\end{equation}
where without the loss of generality the maximum is assumed at
$\phi=0$ . Then, the probability density to find a certain field
value $\phi$ has form \cite{bib:Khlopov1} (adapted to the
considered case):
\begin{multline}\label{prob}
 dP(\phi,T;\phi_{\rm{in}},0) = d\phi\sqrt{\frac{a}{\pi(e^{2\mu T}-1) }} \times\\
 \times\exp\left[-a\frac{(\phi-\phi_{\rm{in}}e^{\mu T})^{2}}{e^{2\mu T}-1}\right].
\end{multline}
Here $a=\mu/\sigma^{2}$, $\mu\equiv m^{2}/3H$ and $\sigma=
H^{3/2}/2\pi$, where the Hubble parameter $H\simeq\sqrt{8\pi
V_{0}/(3M_{\rm{Pl}})}$.

Let us choose a positive value for the initial field,
$\phi_{\rm{in}}>0$. Then an average field value will increase with
time, ultimately reaching the minimum of the potential at some
value $\phi_{+}>0$. This means that a greater part of space will
be finally filled with the field value $\phi=\phi_{+}$. Meanwhile,
the field in some (small) space domain could jump with the
probability (\ref{prob}) over the maximum due to the quantum
fluctuations. In the following, an average value of the field
representing this fluctuation tends to an another minimum of the
potential, $\phi_{-}<0$. As the result, space at the final stage
will be filled by vacuum $\phi_{+}$ while some space domain is
characterized by the field value $\phi=\phi_{-}< 0$. If one starts
to move from inside of the domain to the outside, the path would
start from a space point with $\phi_{-}$ and finish at a space
point with $\phi_{+}$. Hence, the path must contain the point with
the maximum value of potential. It means that a wall is formed
inevitably between such space domains and the ``outer'' space with
$\phi=\phi_{+}$ \cite{bib:Khlopov1,PBH}.

The ``dangerous'' values of fluctuations are those with $\phi\leq
0$. Such space domains will be surrounded by closed walls and if
their number is sufficiently large it would strongly influence the
dynamics of the early Universe. If a fraction of space surrounded
by the walls is not very large, the resulting massive BHs, which
are formed from the walls, could explain the early formation of
quasars \cite{DER}.

\section{Specific features of black hole clusters}\label{sec:clusters}

We will be founded on the described above scenario of primordial
black hole formation, see also \cite{bib:Dokuchaev2,
bib:Dokuchaev4, bib:Rubin1, Ru01b, Ru05, bib:Khlopov1}. This
scenario predicts the existence of supermassive black holes in
galactic nuclei and the existence of  intermediate-mass black
holes in galactic halos at large distances from their centers. One
of the feature of this scenario is the formation of black holes
with a characteristic cluster structure.

The scenario of the formation of massive primordial black holes cannot predict their present mass distribution, since it strongly depends on initial conditions in the period of inflation and the parameters of initial Lagrangian. Moreover, the initial distribution of primordial black holes is distorted during the subsequent evolution of a galaxy when clusters of black holes are merged with each other and with the supermassive black hole in the galactic center. The remaining primordial black holes form a population of black holes in the galactic halo. Thus, for our aims, it is sufficient to determine the initial conditions under
which the number of formed primordial black holes is certainly larger than the number of the unidentified gamma sources.

\begin{figure}
  \includegraphics[width=9cm]{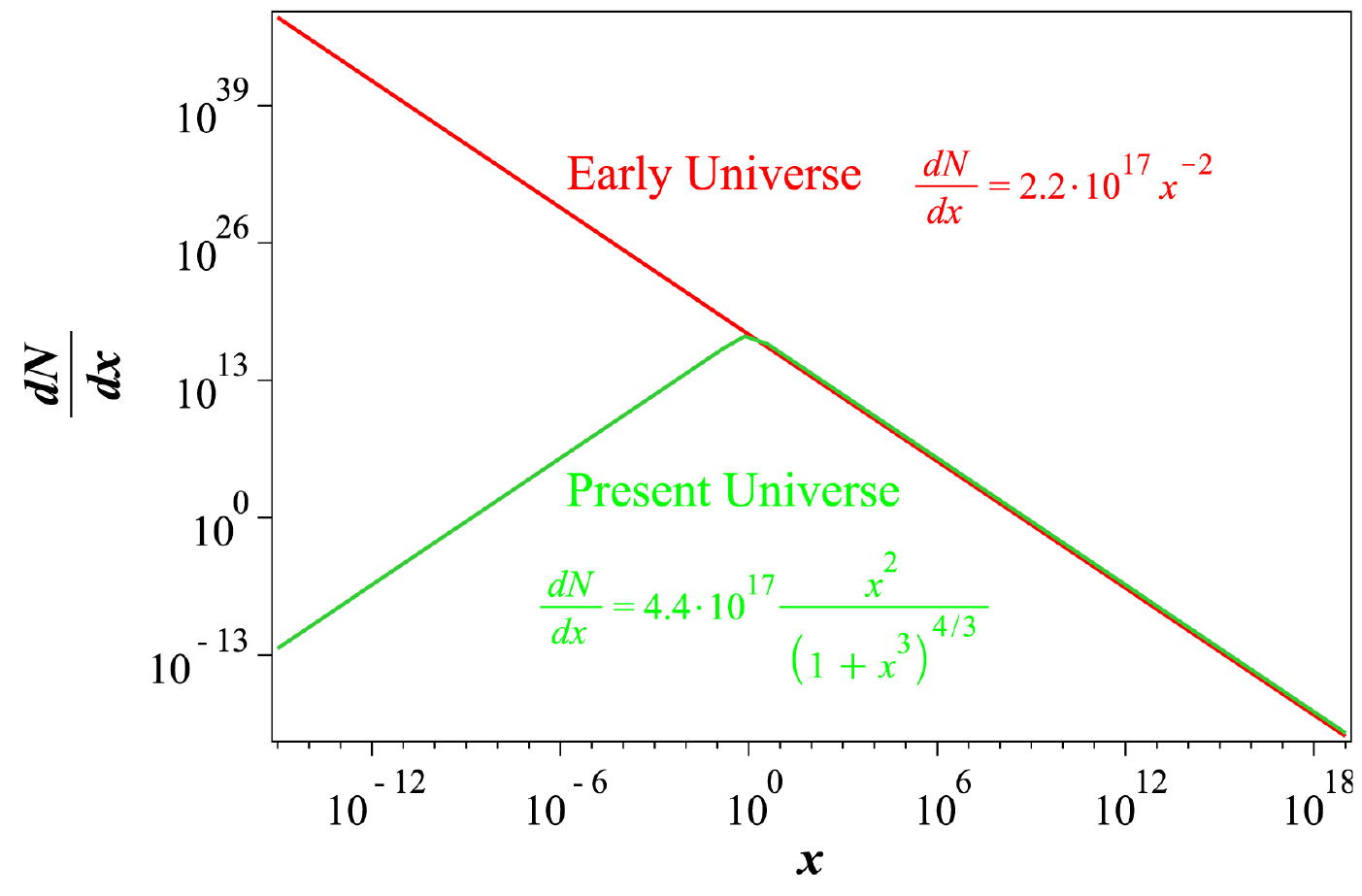}\\
  \caption{Mass distribution of black holes in a cluster. $x = M/M_*$.}
  \label{fig:BHSpectrum}
\end{figure}

The procedure for determining the mass spectrum  of primordial black holes was considered, e.g., in \cite{bib:Dokuchaev1, bib:Dokuchaev2}, where the potential
\begin{equation}\label{MexicLagr}
    V(|\phi|,\theta )=\lambda \left( \phi^* \phi -f^2 /2\right)^2 +
    \Lambda^4 \left( 1 -\cos \theta \right)
\end{equation}
was chosen for certainty. It was shown that at the reasonable
magnitudes of parameters $\Lambda$, $f$ and $\lambda$ ($f = 10.0$
and $\Lambda = 1.66$ in the units of Hubble parameter at the
inflationary stage and almost arbitrary $\lambda$), one obtains an
appropriate structure of clusters.  PBHs within a cluster have
mass distribution approximated by the power law (see
Fig.~\ref{fig:BHSpectrum}) with typical total mass $\sim 10
M_{\odot}$. \beq \frac{dN}{dM}  = f_{\rm{in}}(M_{\rm{in}}) =
\frac{2.2\times10^{17}}{M_*}
\left(\frac{M_*}{M_{\rm{in}}}\right)^{2}.
    \label{eq:f}
\eeq Black holes with masses $M_{\rm{in}} \sim M_* \sim 10^{15}$ g
are of particular interest, because they exist at the final, most
intense evaporation stage now and produce gamma-ray fluxes.
Moreover, observational constraints on them are particularly
strong \cite{bib:Carr}.

As was mentioned above, the resulting mass spectrum is strongly
distorted due to merging of black holes, accretion, and
evaporation. The last process is important for low-mass black
holes with masses $M < M_* \sim 10^{15}$g. Note that non-vanishing
mass distribution in this region is supported by Hawking
evaporation of bigger black holes within the same cluster. Let us
estimate the present evaporation-induced distortion of the mass
spectrum of primordial black holes.

The black-hole evaporation rate is characterized by the temperature \cite{bib:Hawking}
\beq
    T = \frac{1}{8\pi} \frac{\mpl^2}{M}=21\frac{M_*}{M}\;\MeV,
    \label{eq:T}
\eeq
where $\mpl = 1.2\times 10^{19}$ GeV is the Planck mass.

In the approximation of high radiation energies, the effective
evaporation area of black holes is independent of the energy of
the particles \beq
    \sigma= 27\pi M^2/\mpl^4.
\eeq
As a result, the black hole mass loss rate is given by the simple expression
\beq
    \frac{dM}{dt} = -\kappa\frac{dE_{\gamma}}{dt} = -\kappa\frac{\rm const}{M^2},
    \label{eq:dM/dt}
\eeq
where the coefficient $\kappa$ presents the contributions from all of the particles divided by the contribution from photons. Neglecting the dependence of $\kappa$ on $M$ and using Eq.~\eqref{eq:dM/dt}, we obtain the known relation between the initial and present values of the mass of black holes:
\beq
    M_{\rm{in}} = \left( M^3 + M_*^3 \right)^{1/3}.
    \label{eq:M}
\eeq
In this case, the mass distribution of black holes in the cluster given by Eq.~\eqref{eq:f} is transformed as (see Fig.~\ref{fig:BHSpectrum})
\begin{multline}\label{eq:dN/dM}
 f(M) = \frac{dM_{\rm{in}}}{dM}f_{\rm{in}}(M_{\rm{in}}(M))= \\
      = \frac{4.4\times10^{17}}{M_*} \frac{\left(M/M_*\right)^2}{\left(1 + \left(M/M_*\right)^3\right)^{4/3}}.
\end{multline}

The predicted total number of black holes in the cluster
and its mass in the present Universe are $N=\int f(M) dM \sim
4\times10^{17}$ and $M_{\rm{tot}} \sim \int M f(M) dM \sim 9.5
M_\odot$, respectively.

The initial conditions of the formation of black holes in the
simulation are chosen in such way that the number of the formed
clusters of black holes is certainly larger than the number of
gamma-ray sources. Under the chosen conditions, the galactic halo
in the early Universe contained about $N_{\rm{cl}} \sim 1400$
clusters with the typical sizes $R\lesssim 1~\pc$.

The lifetime of such a cluster due to black hole escape can be estimated as \cite{bib:Dokuchaev2,bib:Spitzer}
\beq
    t \approx 40 t_{\rm{rel}},
\eeq
where $t_{\rm{rel}}$ is the relaxation time \cite{bib:Dokuchaev2, bib:Spitzer}
\beq
    t_{\rm{rel}} \approx \frac{1}{4\pi}\frac{v^3}{G^2 m^2 n \log{0.4 N}},
\eeq
$v \sim \sqrt{GmN/R} \sim 3\times10^3 ~\cm/\s$ is the velocity
of black holes in the cluster and $m$, $N$, and $n \sim N/R^3$ are mass, number, and density of typical black holes in the cluster respectively. At the end of this interval the gravitational collapse takes place. Detailed discussion applied to this issue is given in \cite{bib:Dokuchaev2}. The lifetime appears to be many orders of magnitude larger than the age of the Universe; therefore, it can be thought that the total number of clusters survive to the present epoch. Despite a considerably low mass of the cluster, its lifetime is long because of the high density of black holes in the cluster. However, no external destructive influences were taken into account so the fraction of survived clusters is quite uncertain parameter.

The strongest limit on the density of PBH  relates to mass
interval around $\sim 10^{15}$ g \cite{bib:Carr}. The density in the present Universe $\Omega_{\rm{PBH}} \sim 3\times 10^{-10}$, obtained in our calculation, does not contradict this limit.

\section{Photon signal from clusters of black holes}\label{sec:flux}

Let us estimate the luminosity of the typical cluster of PBHs. The
intensity of the direct emission of photons from one black hole
with the temperature $T$ is given by the expression
\cite{bib:Carr} \beq
    dN_{\gamma} = \frac{\sigma}{2\pi^2}\frac{E^2}{\exp{(E/T)}-1}\;dE\;dt
    \label{eq:dN}
\eeq
and strongly depends on the mass of the black hole.
The convolution of spectrum \eqref{eq:dN} and mass distribution \eqref{eq:dN/dM} gives the rate of photon emission from the cluster of black holes (at $E>100$ MeV)
\beq
    \dot{N}\approx 6.6\times10^{36}\ \s^{-1}.
    \label{eq:dN/dt}
\eeq
The total luminosity of the cluster of black holes is calculated as
\beq
    L_{\rm{cl}}  \approx \int\limits_{\mpl}^{M_{\rm{max}}} L(M) \frac{dN}{dM}\;dM
            \approx 9.7\times 10^{35}\ \GeV/\s ,
    \label{eq:Lcl}
\eeq
where the luminosity of single black hole is taken into account in the form \cite{bib:Hawking}
\begin{multline}\label{eq:L}
 L(M) = \frac{1}{15360}\frac{\mpl^4}{M^2} =\\
  =2.2\times10^{18}\left(\frac{M_*}{M}\right)^2\ \GeV/\s.
\end{multline}
Note that the mean energy of emitted photons is $\bar{E} \sim L_{cl}/\dot{N} \approx 15\ \MeV$ and, hence, consideration of clusters of black holes as gamma-ray sources is justified.

\begin{figure}
  \includegraphics[width=9cm]{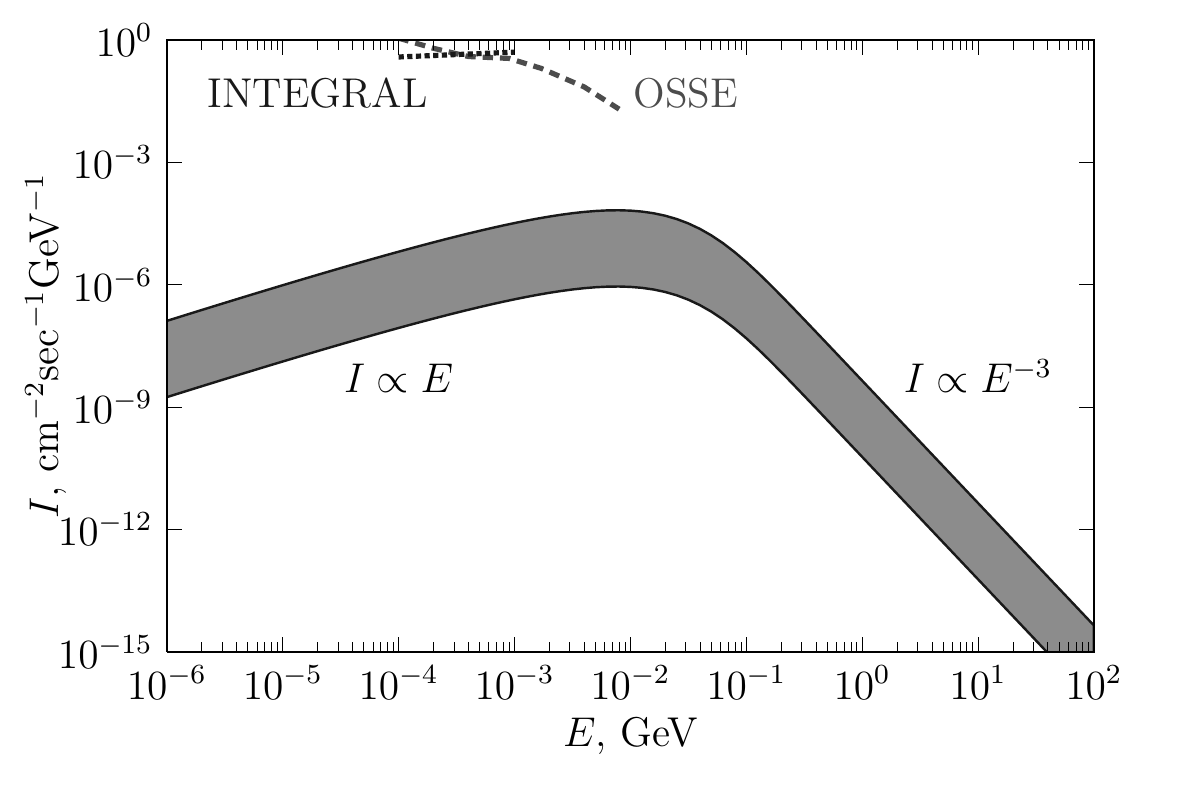}\\
  \caption{The expected photon spectra from PBH clusters (the "banded" region).
Below the region the flux is under LAT integral sensitivity ($r>R_{\rm max}$),
and the expected number of clusters is less than unity above
the region. The differential sensitivity of detectors INTEGRAL and OSSE are also shown.}
  \label{fig:ClSpect}
\end{figure}

Let us determine the number of clusters of PBHs considered as gamma-ray sources that can be observed on the LAT. The photon flux arriving at the Earth from such a cluster is
\beq
    F_{\gamma}  = \frac{\dot{N}}{4\pi r^2},
    \label{eq:F}
\eeq which must exceed, in order to be detected by Fermi LAT, the
threshold value \cite{bib:FERMI.Mission} \beq
    F_{\gamma}^{\rm{min}} = 3\times10^{-9}\ \cm^{-2}\s^{-1}.
    \label{eq:Fmin}
\eeq
Then, the maximum distance $R_{\rm{max}}$ at which the radiation source can be resolved is
\beq
    R_{\rm{max}} = \left(\frac{1}{4\pi}\frac{\dot{N}}{F_{\rm{min}}}\right)^{1/2}
       = 4.3\times10^3\ \pc
    \label{eq:Rmax}
\eeq and therefore the number of such sources is \beq
    N \sim n_{\rm{cl}}\times\frac{4}{3}\pi R_{\rm{max}}^3
      \sim \left(\frac{R_{\rm{max}}}{R_{\rm{gal}}}\right)^3 N_{\rm{cl}}
      \sim 33,
    \label{eq:N}
\eeq
where $n_{\rm{cl}}$ is the number density of clusters, which are
assumed to be located within the characteristic size of the galaxy $R_{\rm{gal}} \sim 15 $~kpc.

The gamma-spectrum seems to be a promising tool for PBH cluster identification.
For energy of interest ($E>100$~MeV) Eq.~\eqref{eq:dN/dM} and Eq.~\eqref{eq:dN} give
\beq
    I = \frac{dN}{dE dt} \propto E^{-3}.
\eeq
In the fig.~\ref{fig:ClSpect} we present the fluxes from PBH clusters.
Note that the maximum at $E\sim10$~MeV in the predicted spectrum corresponds to the peak of mass spectrum
fig.~\ref{fig:BHSpectrum} at $M_\odot\sim10^{15}$~g.

As seen from fig.~\ref{fig:ClSpect} X-ray telescopes are unable to detect PBH clusters because their sensitivity is much less than the threshold value~\cite{bib:INTEGRAL,bib:OSSE}.

\begin{figure}
  \includegraphics[width=9cm]{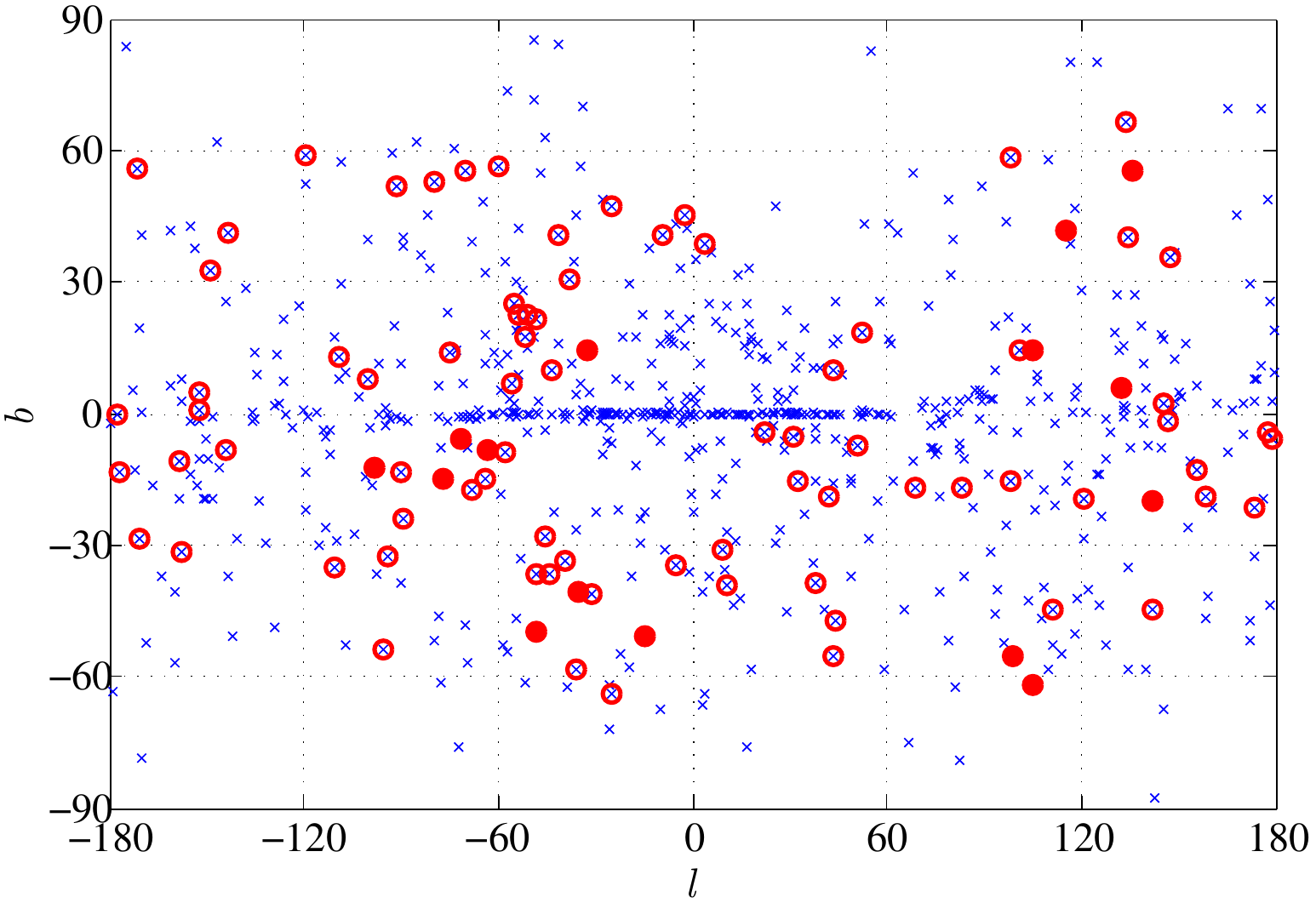}\\
  \caption{Unidentified Gamma ray sources seen by Fermi LAT (blue cross) on celestial sphere are shown. The red filled markers are sources with spectrum indices 3 within $1\sigma$ error, unfilled markers are sources with spectrum indices 3 within $3\sigma$ error.}
  \label{fig:Map}
\end{figure}

The Fermi LAT has detected 15 sources with spectral index $-3$ at $1\sigma$ and 93 sources
at $3\sigma$ error~\cite{bib:1FGL}. Their distribution on the celestial sphere is isotropic and their amount is in good agreement with the prediction of Eq.~\eqref{eq:N} (see fig.~\ref{fig:Map}).

Each PBH cluster should be also a point-like source of high-energy neutrinos. We estimated the expected neutrino flux in similar manner and compared it with sensitivity of neutrino observatories. It was obtained that neutrino observatory AMANDA \cite{bib:AMANDA, bib:SK}, which is sensitive to the flux of neutrinos with energy $>1.9$~GeV exceeding $\sim 10^{-11} \cm^{-1}\s^{-1}$, can potentially detect neutrinos from the cluster at several kpc. At these distances we can expect about one cluster. It means that neutrino experiments (AMANDA, IceCube) are promising for indication of PBH clusters.

\section{Conclusions}\label{sec:concl}

In this work, we attempted to unify two astrophysical problems.
First, we propose the explanation of the origin of unidentified
gamma-ray sources. Second, new tool of low-mass black holes
searching is discussed. Multiple production of the latter is
predicted by some models of the formation of massive primordial black holes in galactic centers. It has been shown that if the spatial distribution of PBHs has a cluster structure, an individual cluster of PBHs is detected as a point-like gamma-source. Total luminosity of small black holes in a distant cluster is large enough to be detected on the Earth. Note that the model of the formation of massive primordial black holes proposed in \cite{bib:Rubin1, Ru01b} predicts just the cluster structure. The mass spectrum of PBHs, their abundance, and their radiation energy spectrum calculated using this model indicate the real possibility of the correlation between point-like gamma-ray sources and PBHs.

\section{Acknowledgments}
We are grateful to V.I. Dokuchaev for valuable remarks. The work of K.B., A.K. and S.R. was supported by The Ministry of education and science of Russia, projects 14.A18.21.0789, 14.132.21.1446 and 8525.

\bibliographystyle{elsarticle-num}
\bibliography{Article}

\begin{thebibliography}{10}
\expandafter\ifx\csname url\endcsname\relax
  \def\url#1{\texttt{#1}}\fi
\expandafter\ifx\csname urlprefix\endcsname\relax\def\urlprefix{URL }\fi
\expandafter\ifx\csname href\endcsname\relax
  \def\href#1#2{#2} \def\path#1{#1}\fi

\bibitem{bib:Dolgov1}
A.~D. Dolgov, J.~Silk, {Baryon isocurvature fluctuations at small scales and
  baryonic dark matter}, Phys. Rev. D47 (1993) 4244--4255.

\bibitem{bib:Dokuchaev1}
V.~I. Dokuchaev, {Birth and Life of Massive Black Holes}, Sov. Phys. Usp. 34
  (1991) 447.

\bibitem{bib:Dolgov2}
A.~D. Dolgov, M.~Kawasaki, N.~Kevlishvili, {Inhomogeneous baryogenesis, cosmic
  antimatter, and dark matter}, Nucl. Phys. B807 (2009) 229--250.
\newblock \href {http://arxiv.org/abs/hep-ph/0806.2986}
  {\path{arXiv:hep-ph/0806.2986}}.

\bibitem{bib:Rees}
M.~J. Rees, {Black Hole Models for Active Galactic Nuclei}, Ann. Rev. Astron.
  Astrophys. 22 (1984) 471--506.

\bibitem{bib:Begelman}
M.~C. Begelman, R.~D. Blandford, M.~J. Rees, {Theory of extragalactic radio
  sources}, Rev. Mod. Phys. 56 (1984) 255--351.

\bibitem{bib:Dokuchaev2}
V.~I. Dokuchaev, Y.~N. Eroshenko, R.~S. G., {Early Formation of Galaxies
  Initiated by Clusters of Primordial Black Holes}, Astron. Rep. 52 (2007)
  779--789.
\newblock \href {http://arxiv.org/abs/astro-ph/0801.0885}
  {\path{arXiv:astro-ph/0801.0885}}.

\bibitem{bib:Dokuchaev3}
V.~I. Dokuchaev, Y.~N. Eroshenko, {Stochastic Correlation Model of Galactic
  Bulge Velocity Dispersions and Central Black Holes Masses}, Astron. Lett. 27
  (2001) 759--764.
\newblock \href {http://arxiv.org/abs/astro-ph/0202019}
  {\path{arXiv:astro-ph/0202019}}.

\bibitem{bib:Dokuchaev4}
V.~I. Dokuchaev, Y.~N. Eroshenko, {Origin of Correlations between Central Black
  Holes Masses and Galactic Bulge Velocity Dispersions}, Astron. Astrophys.
  Trans. 22 (2003) 727.
\newblock \href {http://arxiv.org/abs/astro-ph/0209324}
  {\path{arXiv:astro-ph/0209324}}.

\bibitem{bib:3EG}
R.~C. Hartman, et~al., {The Third EGRET catalog of high-energy gamma-ray
  sources}, Astrophys. J. Suppl. 123 (1999) 79.

\bibitem{bib:EGR}
J.-M. Casandjian, I.~A. Grenier, A revised catalogue of egret $\gamma$-ray
  sources, Astron. \& Astrophys. 489~(2) (2008) 849--883.

\bibitem{bib:Thompson}
D.~J. Thompson, {Gamma ray astrophysics: the EGRET results}, Rept. Prog. Phys.
  71 (2008) 116901.
\newblock \href {http://arxiv.org/abs/astro-ph/0811.0738}
  {\path{arXiv:astro-ph/0811.0738}}.

\bibitem{bib:EGRET.Calibration}
J.~A. Esposito, et~al., In-flight calibration of egret on the compton gamma-ray
  observatory, Astrophys. J. Suppl. 123~(1)  203.

\bibitem{bib:1FGL}
W.~B. Atwood, et~al., {Fermi Large Area Telescope First Source Catalog},
  Astrophys. J. Suppl. 188 (2010) 405--436.
\newblock \href {http://arxiv.org/abs/astro-ph.HE/1002.2280}
  {\path{arXiv:astro-ph.HE/1002.2280}}.

\bibitem{bib:FERMI.Mission}
W.~B. Atwood, et~al., {The Large Area Telescope on the Fermi Gamma-ray Space
  Telescope Mission}, Astrophys. J. 697 (2009) 1071--1102.
\newblock \href {http://arxiv.org/abs/astro-ph.IM/0902.1089}
  {\path{arXiv:astro-ph.IM/0902.1089}}.

\bibitem{bib:Carr}
B.~J. Carr, K.~Kohri, Y.~Sendouda, J.~Yokoyama, {New cosmological constraints
  on primordial black holes}, Phys. Rev. D81 (2010) 104019.
\newblock \href {http://arxiv.org/abs/astro-ph.CO/0912.5297}
  {\path{arXiv:astro-ph.CO/0912.5297}}.

\bibitem{bib:Rubin1}
S.~G. Rubin, M.~Y. Khlopov, A.~S. Sakharov, {Primordial black holes from
  non-equilibrium second order phase transition}, Grav. Cosmol. S6 (2000)
  51--58.
\newblock \href {http://arxiv.org/abs/hep-ph/0005271}
  {\path{arXiv:hep-ph/0005271}}.

\bibitem{Carr}
B.~J. Carr, {Primordial Black Holes: Do They Exist and Are They Useful?}\href
  {http://arxiv.org/abs/astro-ph/0511743} {\path{arXiv:astro-ph/0511743}}.

\bibitem{carr75}
B.~J. Carr, {The Primordial black hole mass spectrum}, Astrophys. J. 201 (1975)
  1--19.

\bibitem{zeldnov66}
Y.~B. Zel'dovich, I.~D. Novikov, {The Hypothesis of Cores Retarded during
  Expansion and the Hot Cosmological Model}, Sov. Astron. J. 10 (1967)
  602--603.

\bibitem{hawking71}
S.~Hawking, {Gravitationally collapsed objects of very low mass}, Mon. Not.
  Roy. Astron. Soc. 152 (1971) 75.

\bibitem{Khlopov80}
M.~Y. Khlopov, A.~G. Polnarev, {Primordial Black Holes As A Cosmological Test
  Of Grand Unification}, Phys. Lett. B97 (1980) 383--387.

\bibitem{Barrow81}
J.~D. Barrow, M.~S. Turner, {Baryosynthesis And The Origin Of Galaxies}, Nature
  291 (1981) 469--472.

\bibitem{Bond82}
J.~R. Bond, J.~Silk, E.~W. Kolb, {The generation of isothermal perturbations in
  the very early universe}, Astrophys. J. 255 (1982) 341--360.

\bibitem{Fuk86}
M.~Fukugita, V.~A. Rubakov, {A Possible Mechanism Of Generating Almost
  Isothermal Baryon Density Perturbations}, Phys. Rev. Lett. 56 (1986) 988.

\bibitem{Dolg87}
A.~D. Dolgov, A.~F. Illarionov, N.~S. Kardashev, I.~D. Novikov, {A cosmological
  model of the baryon island}, Sov. Phys. JETP 94 (1988) 1--14.

\bibitem{Kof87}
L.~A. Kofman, A.~D. Linde, {Generation of Density Perturbations in the
  Inflationary Cosmology}, Nucl. Phys. B282 (1987) 555.

\bibitem{Kof88}
L.~A. Kofman, D.~Y. Pogosian, {Nonflat Perturbations In Inflationary
  Cosmology}, Phys. Lett. B214 (1988) 508--514.

\bibitem{Dol91}
A.~D. Dolgov, D.~P. Kirilova, {Baryon charge condensate and baryogenesis}, J.
  Moscow Phys. Soc. 1 (1991) 217--229.

\bibitem{CDol92}
M.~V. {Chizhov}, A.~D. {Dolgov}, {Baryogenesis and large-scale structure of the
  universe}, Nuclear Physics B 372 (1992) 521--529.

\bibitem{Iok91}
J.~Yokoyama, Y.~Suto, {Baryon isocurvature scenario in inflationary cosmology:
  A particle physics model and its astrophysical implications}, Astrophys. J.
  379 (1991) 427--439.

\bibitem{DolgovSilk93}
A.~D. Dolgov, J.~Silk, {Baryon isocurvature fluctuations at small scales and
  baryonic dark matter}, Phys. Rev. D47 (1993) 4244--4255.

\bibitem{Star92}
A.~A. Starobinskij, {Spectrum of adiabatic perturbations in the universe when
  there are singularities in the inflationary potential.}, Sov. Phys. JETP 55
  (1992) 489--494.

\bibitem{ivan94}
P.~Ivanov, P.~Naselsky, I.~Novikov, {Inflation and primordial black holes as
  dark matter}, Phys. Rev. D50 (1994) 7173--7178.

\bibitem{Yoko}
J.~Yokoyama, {Chaotic new inflation and formation of primordial black holes},
  Phys. Rev. D58 (1998) 083510.
\newblock \href {http://arxiv.org/abs/astro-ph/9802357}
  {\path{arXiv:astro-ph/9802357}}.

\bibitem{Jedamzik}
K.~Jedamzik, {Primordial black hole formation during the QCD epoch}, Phys. Rev.
  D55 (1997) 5871--5875.
\newblock \href {http://arxiv.org/abs/astro-ph/9605152}
  {\path{arXiv:astro-ph/9605152}}.

\bibitem{Jedamzik2}
K.~Jedamzik, {Could MACHOS be primordial black holes formed during the QCD
  epoch?}, Phys. Rept. 307 (1998) 155--162.
\newblock \href {http://arxiv.org/abs/astro-ph/9805147}
  {\path{arXiv:astro-ph/9805147}}.

\bibitem{Bubble}
M.~Crawford, D.~N. Schramm, {Spontaneous generation of density perturbations in
  the early universe}, Nature 298 (1982) 538--540.

\bibitem{Bubble2}
S.~W. Hawking, I.~G. Moss, J.~M. Stewart, {Bubble Collisions in the Very Early
  Universe}, Phys. Rev. D26 (1982) 2681.

\bibitem{Dodelson}
S.~{Dodelson}, {Modern cosmology}, Academic Press, 525 B Street, Suite 1900,
  San Diego, California 92101-4495, USA, 2003.

\bibitem{Bubble3}
H.~Kodama, M.~Sasaki, K.~Sato, {Abundance Of Primordial Holes Produced By
  Cosmological First Order Phase Transition}, Prog. Theor. Phys. 68 (1982)
  1979.

\bibitem{Bubble4}
D.~La, P.~J. Steinhardt, {Bubble Percolation in Extended Inflationary Models},
  Phys. Lett. B220 (1989) 375.

\bibitem{Bubble5}
I.~G. Moss, {Singularity formation from colliding bubbles}, Phys. Rev. D50
  (1994) 676--681.

\bibitem{Bubble6}
R.~V. Konoplich, S.~G. Rubin, A.~S. Sakharov, M.~Y. Khlopov, {Black hole
  production in first order phase transitions in the Universe}, Sov. Pis'ma
  Astron. J. 24 (1998) 1.

\bibitem{Bubble7}
R.~V. {Konoplich}, S.~G. {Rubin}, A.~S. {Sakharov}, M.~Y. {Khlopov}, {Formation
  of black holes in first-order phase transitions as a cosmological test of
  symmetry-breaking mechanisms}, Phys. Atom. Nucl. 62 (1999) 1593--1600.

\bibitem{Ru25}
I.~{Dymnikova}, L.~{Koziel}, M.~{Khlopov}, S.~{Rubin}, {Quasilumps from First
  Order Phase Transitions}, Grav. \& Cosm. 6 (2000) 311--318.
\newblock \href {http://arxiv.org/abs/hep-th/0010120}
  {\path{arXiv:hep-th/0010120}}.

\bibitem{Ru01b}
S.~G. {Rubin}, A.~S. {Sakharov}, M.~Y. {Khlopov}, {The Formation of Primary
  Galactic Nuclei during Phase Transitions in the Early Universe}, Sov. Phys.
  JETP 92 (2001) 921--929.
\newblock \href {http://arxiv.org/abs/hep-ph/0106187}
  {\path{arXiv:hep-ph/0106187}}.

\bibitem{Ru05}
M.~Y. {Khlopov}, S.~G. {Rubin}, A.~S. {Sakharov}, {Primordial structure of
  massive black hole clusters}, Astropart. Phys. 23 (2005) 265--277.
\newblock \href {http://arxiv.org/abs/astro-ph/0401532}
  {\path{arXiv:astro-ph/0401532}}.

\bibitem{GraCo}
K.~M. {Belotsky}, A.~V. {Berkov}, A.~A. {Kirillov}, S.~G. {Rubin}, {Black hole
  clusters in our Galaxy}, Grav. \& Cosm. 17 (2011) 27--30.

\bibitem{Ru10}
V.~I. {Dokuchaev}, Y.~N. {Eroshenko}, S.~G. {Rubin}, D.~A. {Samarchenko},
  {Mechanism for the suppression of intermediate-mass black holes}, Sov.
  Astron. Lett. 36 (2010) 773--779.
\newblock \href {http://arxiv.org/abs/astro-ph.CO/1010.5325}
  {\path{arXiv:astro-ph.CO/1010.5325}}.

\bibitem{bib:Hawking}
S.~W. Hawking, {Particle Creation by Black Holes}, Commun. Math. Phys. 43
  (1975) 199--220.

\bibitem{Dvali}
G.~Dvali, S.~Kachru, {New old inflation}\href
  {http://arxiv.org/abs/hep-th/0309095} {\path{arXiv:hep-th/0309095}}.

\bibitem{Racetrack}
J.~J. {Blanco-Pillado}, et~al., {Racetrack Inflation}, J. High Energy Phys. 11
  (2004) 63.
\newblock \href {http://arxiv.org/abs/hep-th/0406230}
  {\path{arXiv:hep-th/0406230}}.

\bibitem{LindeHyb}
A.~D. Linde, {Axions in inflationary cosmology}, Phys. Lett. B259 (1991)
  38--47.

\bibitem{Dolgov97}
A.~Dolgov, K.~Freese, R.~Rangarajan, M.~Srednicki, {Baryogenesis during
  reheating in natural inflation and comments on spontaneous baryogenesis},
  Phys. Rev. D56 (1997) 6155--6165.
\newblock \href {http://arxiv.org/abs/hep-ph/9610405}
  {\path{arXiv:hep-ph/9610405}}.

\bibitem{Star}
A.~A. {Starobinsky}, {Stochastic de Sitter (inflationary) Stage in the Early
  Universe}, in: {H.~J.~de Vega \& N.~S{\'a}nchez} (Ed.), Field Theory, Quantum
  Gravity and Strings, Vol. 246 of Lecture Notes in Physics, Berlin Springer
  Verlag, 1986, pp. 107--124.

\bibitem{bib:Khlopov1}
M.~Y. Khlopov, S.~G. Rubin, Cosmological Pattern of Microphysics in the In
  ationary Universe, Kluwer Academic Publishers, P.O. Box 17, 3300 AA
  Dordrecht, The Netherlands, 2004.

\bibitem{PBH}
S.~G. Rubin, {Massive Primordial Black Holes in Hybrid Inflation}\href
  {http://arxiv.org/abs/astro-ph/0511181} {\path{arXiv:astro-ph/0511181}}.

\bibitem{DER}
V.~I. {Dokuchaev}, Y.~N. {Eroshenko}, S.~G. {Rubin}, {Quasars formation around
  clusters of primordial black holes}, Grav. \& Cosm. 11 (2005) 99--104.
\newblock \href {http://arxiv.org/abs/astro-ph/0412418}
  {\path{arXiv:astro-ph/0412418}}.

\bibitem{bib:Spitzer}
J.~L. {Spitzer}, W.~C. {Saslaw}, {On the Evolution of Galactic Nuclei},
  Astrophys. J. 143 (1966) 400--419.

\bibitem{bib:INTEGRAL}
C.~Winkler, et~al., The integral mission, Astron. \& Astrophys. 411 (2003)
  L1--L6.

\bibitem{bib:OSSE}
R.~A. Cameron, et~al., {Operation and performance of the OSSE instrument}, in:
  NASA Conference Publication, Vol. 3137 of NASA Conference Publication, 1992.

\bibitem{bib:AMANDA}
R.~Abbasi, et~al., {Search for Point Sources of High Energy Neutrinos with
  Final Data from AMANDA-II}, Phys. Rev. D79 (2009) 062001.
\newblock \href {http://arxiv.org/abs/astro-ph/0809.1646}
  {\path{arXiv:astro-ph/0809.1646}}.

\bibitem{bib:SK}
E.~Thrane, et~al., {Search for Astrophysical Neutrino Point Sources at Super-
  Kamiokande}, Astrophys. J. 704 (2009) 503--512.
\newblock \href {http://arxiv.org/abs/astro-ph.HE/0907.1594}
  {\path{arXiv:astro-ph.HE/0907.1594}}.

\end{thebibliography}

\end{document}